\documentclass{kluwer}    

\usepackage{psfig}

\def\etal{{et\,al. }}
\def\degs{\ifmmode ^{\circ}\else$^{\circ}$\fi}
\newbox\grsign \setbox\grsign=\hbox{$>$}
\newdimen\grdimen \grdimen=\ht\grsign
\newbox\laxbox \newbox\gaxbox
\setbox\gaxbox=\hbox{\raise.5ex\hbox{$>$}\llap
     {\lower.5ex\hbox{$\sim$}}}\ht1=\grdimen\dp1=0pt
\setbox\laxbox=\hbox{\raise.5ex\hbox{$<$}\llap
     {\lower.5ex\hbox{$\sim$}}}\ht2=\grdimen\dp2=0pt

\begin{document}                                                                                   
\begin{article}
\begin{opening}         
\title{Broadband X-ray Spectroscopy of GRS 1915+105
} 
\author{A. \surname{Rau}, J. \surname{Greiner}}  
\runningauthor{Rau \& Greiner}
\runningtitle{Broadband X-ray Spectroscopy of GRS 1915+105}
\institute{Astrophysical Institute Potsdam, 14482 Potsdam, Germany}


\begin{abstract}
We analyzed RXTE data of GRS 1915+105 of X-ray low states between 1996 and 1998. The X-ray spectrum is dominated by the power law component. We found (i) that the power law is pivoting at 10-30 keV, (ii) that the power law slope correlates with radio flux, (iii) three different groups  of 3-10~keV residuals, (iv) that one of these residual groups correlates with the power law slope.
\end{abstract}


\end{opening}           
\bigskip

The prototypical microquasar GRS 1915+105 was discovered by Granat (Castro-Tirado, Brandt \& Lund 1992) as a transient X-ray source. Observations with RXTE PCA revealed astonishing X-ray variability (Greiner \etal\ 1996) which has been classified into a dozen different X-ray states (Belloni \etal\ 2000).


We analyzed 68 observations from the RXTE public archive of GRS 1915+105 during X-ray low states, so called \(\chi\)-states (Belloni \etal\ 2000; no large amplitude variations nor obvious structured variability) from November 1996 to June 1998. For each observation we used PCU 0 data of the PCA and binned \mbox{HEXTE} cluster A data and fitted these with a model consisting of cold absorption (wabs), disk  blackbody (diskbb) and reflection (refsch, including a power law) in \em XSPEC \rm with a fixed folding energy of 400~keV and solar abundances. Energies between 4 and 8.5~keV were ignored during the fit.

All reduced X-ray spectra can be divided into three groups depending on the residuals in the energy range 3-10~keV. The first group shows a small hump at 6.4~keV and a dip at 4.5~keV, and the second group has two broad humps at 4.5 and 6~keV. Both groups are mainly  seen during the long 1997 low state. The third group has a deep dip at 4.5~keV as the only feature and occurs only at times of radio flares.
A deepening of the ``dip''-residuals with steeper power law and increasing radio flux is seen, whereas no such trend is visible for the other groups.
This correlation between dip depth and radio flares suggests that the depth is connected with the strengths of a possibly nearly-spherical outflow or accretion disk wind (inclination is 70\degs~from radio jet characteristics, e.g. nearly edge-on) where a part may be collimated into a jet.

 \begin{figure}[th]
   \centering{
   \hspace{-0.01cm}
   \vbox{\psfig{figure=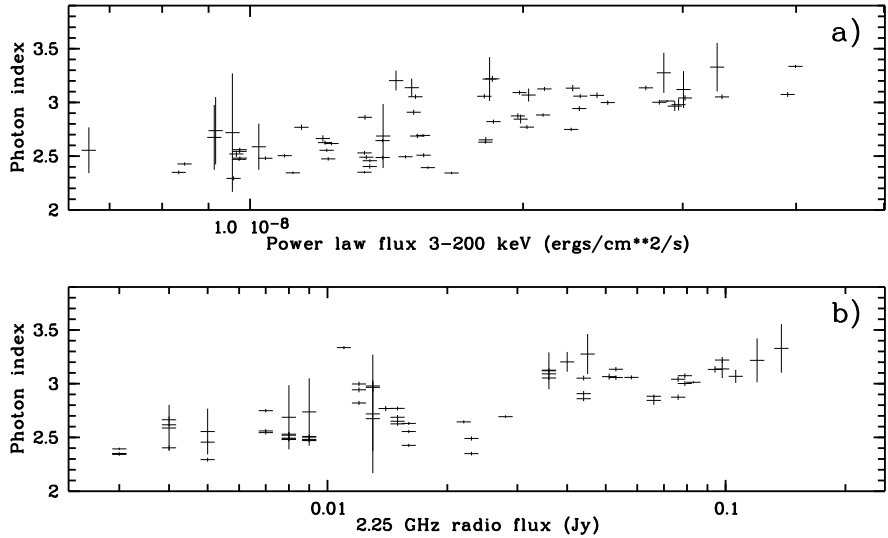,width=1.\textwidth,bbllx=2.8cm,bblly=15.5cm,bburx=18.7cm,bbury=26.cm,clip=}}\par
   \vspace{-0.25cm}
   \caption[lc]{X-ray power law photon index as function of a) the 3-200~keV power law flux and b) the 2.25~GHz Green Bank Interferometer (GBI) radio flux.
          }}
         \label{lc}
 \end{figure}

\bigskip
During \(\chi\)-states (independent of the three residual groups) an increasing PCA count rate is related to a decreasing HEXTE count rate and an increasing spectral photon index.
Interestingly, the power law pi\-votes at 10-30 keV, while the disk component is always negligible. Therefore a factor of 3-4 in PCA count rate can be explained by a varying power law flux by the same factor (Fig.~1a). Thus, the power law producing corona of hot electrons can be the origin of the observed count rate variation. An increasing photon index is caused by a decreasing electron temperature in the corona and the observed variation implies a temperature variation between 25 and 50 keV (Sazonov \& Sunyaev 2000).

Also correlated with the power law slope is the 2.25~GHz radio flux as measured with the GBI such that a higher photon index (lower electron temperature) is related to higher radio flux (Fig.~1b). This could be due to a shift of the synchrotron radiation to longer wavelengths with decreasing electron temperature. This would be in agreement to a correlation of the K band flux with the X-ray hardness ratio (Greiner et al. 2001).

\end{article}

\begin{thebibliography}{}

\bibitem[\protect\citeauthoryear{}{}]{}
Belloni T., Klein-Wolt M., Mendez M. \etal\ 
2000, {\it A\&A} 355, 271
\bibitem[\protect\citeauthoryear{}{}]{}
Castro-Tirado A.J., Brandt S., Lund N.,  
1992, {\it IAU Circ.}, 5590 
\bibitem[\protect\citeauthoryear{}{}]{}
Greiner J., Morgan E., Remillard R.A., 
1996, {\it ApJ}, 473, L107
\bibitem[\protect\citeauthoryear{}{}]{}
Greiner J., Vrba F.J., Henden A.A. \etal\ 
2001, these proceedings
\bibitem[\protect\citeauthoryear{}{}]{}
Sazonov S.Y., Sunyaev R.A., 
2000, {\it AstL}, 26, 494


\end{thebibliography}
\end{document}